\documentclass[letterpaper,twocolumn,10pt]{article}
\usepackage{usenix2019_v3}

\usepackage{tikz}
\usepackage{amsmath}

\usepackage{booktabs}       
\usepackage{algorithm}
\usepackage{algpseudocode}
\usepackage{multirow}
\usepackage{amsfonts}
\usepackage{amsmath}
\usepackage{amssymb}
\usepackage[utf8]{inputenc} 
\usepackage{nicefrac}       
\usepackage{subcaption}
\usepackage{float}

\newcommand{\Z}{{\mathbb{Z}}}
\newcommand{\C}{{\mathbb{C}}}

\newcommand{\Mod}[1]{\ (\mathrm{mod}\ #1)}

\newcommand{\privft}{PrivFT}
\newcommand{\ahe}{A\(^\ast\)HE}
\newcommand{\fasttext}{\textsf{fasttext}}

\begin{document}

\date{}

\title{\privft{}: Private and Fast Text Classification with Homomorphic Encryption}

\author{
{\rm Ahmad Al Badawi}\\
I$^2$R, A*STAR\\
{\tt \footnotesize{Ahmad\_Al\_Badawi@i2r.a-star.edu.sg}}
\and
{\rm Luong Hoang}\\
Inferati Inc.\\
{\tt \footnotesize{luong@inferati.com}}
\and
\rm{Chan Fook Mun}\\
I$^2$R, A*STAR\\
{\tt \footnotesize{Chan\_Fook\_Mun@i2r.a-star.edu.sg}}
\and
{\rm Kim Laine}\\
Microsoft Research\\
{\tt \footnotesize{kim.laine@microsoft.com}}
\and
\rm{Khin Mi Mi Aung}\\
I$^2$R, A*STAR\\
{\tt \footnotesize{Mi\_Mi\_Aung@i2r.a-star.edu.sg}}
} 

\maketitle

\begin{abstract}

The need for privacy-preserving analytics is higher than ever due to the severity of privacy risks and to comply with new privacy regulations leading to an amplified interest in privacy-preserving techniques that try to balance between privacy and utility. In this work, we present an efficient method for Text Classification while preserving the privacy of the content using Fully Homomorphic Encryption (FHE). Our system (named \textbf{Priv}ate \textbf{F}ast \textbf{T}ext (PrivFT)) performs two tasks: 1) making inference of encrypted user inputs using a plaintext model and 2) training an effective model using an encrypted dataset. For inference, we train a supervised model and outline a system for homomorphic inference on encrypted user inputs with zero loss to prediction accuracy. In the second part, we show how to train a model using fully encrypted data to generate an encrypted model. We provide a GPU implementation of the Cheon-Kim-Kim-Song (CKKS) FHE scheme and compare it with existing CPU implementations to achieve 1 to 2 orders of magnitude speedup at various parameter settings. We implement PrivFT in GPUs to achieve a run time per inference of less than 0.66 seconds. Training on a relatively large encrypted dataset is more computationally intensive requiring 5.04 days.
\end{abstract}

\section{Introduction}\label{sec:introduction}


Machine Learning (ML) has become widely adopted in critical electronic systems that make high-stakes decisions in several domains, such as healthcare, law and finance~\cite{chen2017disease,mena2016machine,heaton2017deep}. These systems deal usually with private data collected from data owners (individuals) or data stores, such as hospitals, financial institutions and civil departments. ML models used in such systems are usually created (trained) and used (make inference) in-house to guarantee that data privacy and confidentiality are not compromised. As ML models are becoming increasingly more complex due to the need for higher prediction accuracy, a dedicated team of ML and domain experts should be employed by different institutions to generate and maintain these models incurring high additional cost. 

An alternative solution is to leverage the capabilities of the cloud and lease the usage of pre-learned models hosted by the cloud in a service model known as ML as a Service (MLaaS). In this scenario, the client requests the cloud to evaluate a model on her inputs and return the inference result. Two problems arise in this scenario: 1) how to generate the model on the cloud, and 2) privacy and confidentiality of the clients' input might be compromised by the cloud at inference. Even if the cloud can be assumed to be trusted, in such critical application domains, sharing private data with a third-party might be considered against the law. 

In order to navigate these conflicting objectives, researchers have been looking into privacy-preserving techniques that can balance between privacy and utility. One promising technique is the Fully Homomorphic Encryption (FHE), which is a new class of encryption schemes that allow computing on encrypted data without decryption~\cite{STOC:Gentry09}. The key assumption here is that as long as the data is encrypted with a secure cryptographic scheme using a private key that only intended parties have, encrypted data can be released in public. FHE encrypted data can be operated on by an untrusted evaluator to generate encrypted results without revealing intermediate/final results to the evaluator. The results are communicated back to the data owner who can decrypt and make use of them.

FHE has been shown to be useful in a wide range of application domains such as image classification~\cite{MSFT:DGL+16,badawi2018alexnet,chou2018faster}, statistical analysis~\cite{ARXIV:AslEspHol15,cryptoeprint:2016:1163}, and Genome-Wide Association Studies (GWAS)~\cite{wang2015healer,lu2015privacy}. Despite its incredible capabilities, FHE suffers from two major issues: 1) high computational overhead and 2) limited arithmetic set (only addition and multiplication on encrypted data are supported). The latter means that one has to build the desired function as a circuit\footnote{Circuits are built using $\Mod{p}$ gates, where $p \geq 2 \in \Z$, where $\Z$ is the set of integers. In the case where $p = 2$, these are the normal binary circuits. When $p>2$, circuits are composed of arithmetic gates with operands and outputs $\in \Z_p$.} so it can be evaluated with FHE.

In this paper, we target Text Classification as a new application domain for homomorphic evaluation of both training and inference on encrypted data. As a baseline model, we choose a shallow artificial neural network (\fasttext{})~\cite{joulin2016bag}, proposed for the task of text classification. 
Using simple techniques, the model achieves competitive results to those with more complex architectures. 
More importantly, this choice allows us to build an FHE-friendly solution that can perform classification directly over encrypted text data. 

\subsection{System Overview}
The system proposed in this paper, which we call \textbf{Priv}ate \textbf{F}ast \textbf{T}ext (\privft), performs two main tasks as described below:
\begin{itemize}
    \item Homomorphic inference on encrypted data: in this task, we focus on how to perform inference on encrypted input texts. We consider a secure (MLaaS) system where a Natural Language Processing (NLP) model, previously trained on non-encrypted data, is stored in the cloud in plaintext form. The client, or data owner, uses a homomorphic encryption scheme to transform her plaintext input into encrypted form (ciphertext) and sends it to the cloud model for inference. This evaluation generates an encrypted prediction of the class the text belongs to that is sent back to the client who can decrypt and obtain the output in plaintext form. As we shall see later, this task is very efficient with computation time less than 0.66 seconds making it highly practical in modern applications.

    \item Homomorphic training on encrypted data: in this task, we focus on how to train a model from scratch using encrypted dataset. Here, the data owner sends her encrypted data, using an FHE scheme, to the cloud which in turn performs a batched training algorithm using back-propagation to learn an encrypted model (i.e. training operations are done entirely in ciphertext space). The encrypted model is sent back to the client who can decrypt and use for local inference. Unlike the previous inference as a service, this task requires high computational demands. As training is usually performed less frequently, the task can still be used in some application domains.
\end{itemize}

We emphasize that in both tasks no decryption takes place at the cloud side, which makes our solution as secure as the encryption scheme itself. To the best of our knowledge, the homomorphic encryption scheme used here is provably secure when the parameters are set appropriately.

\begin{figure*}[!ht]
    \centering
    \begin{subfigure}[t]{0.55\textwidth}
        \centering
        \includegraphics[height=4.7cm, width=1\textwidth]{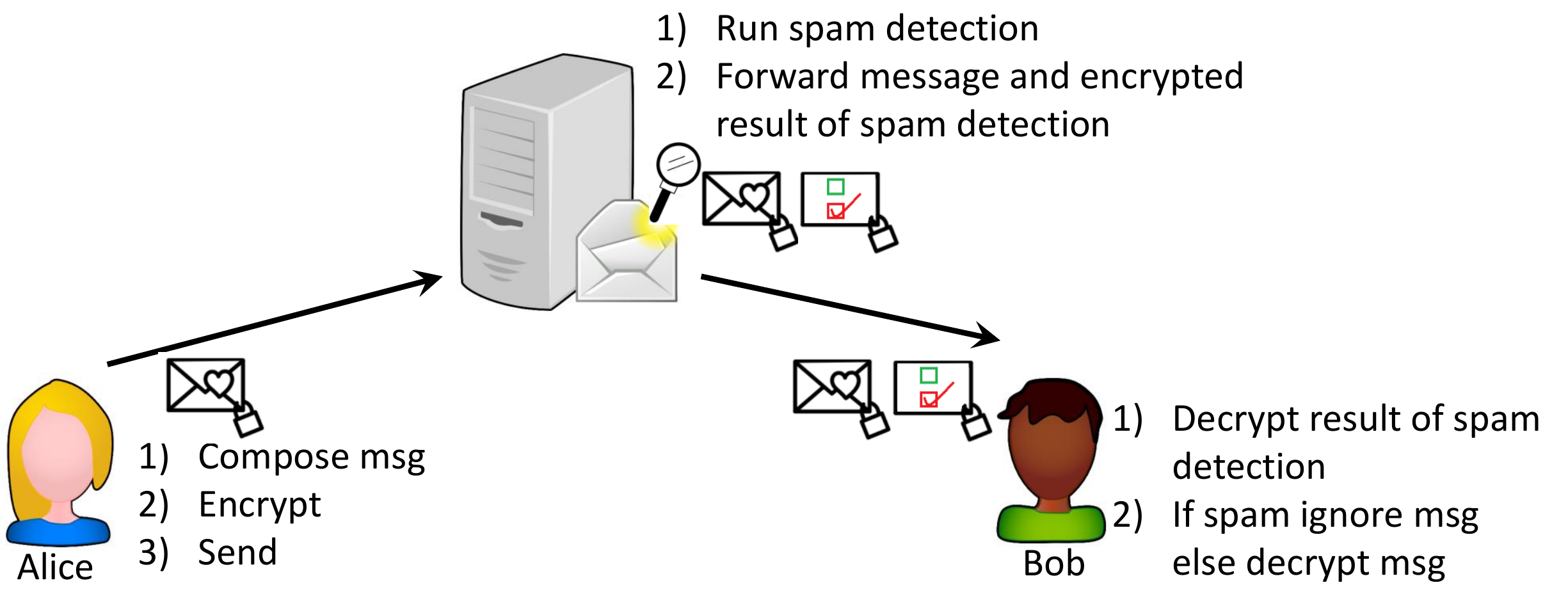}
        \caption{Fully encrypted e-mail service with enabled spam detection via inference as a service.}
        \label{fig:inference:as:a:service}
    \end{subfigure}%
    ~~~
    \begin{subfigure}[t]{0.45\textwidth}
        \centering
        \includegraphics[height=4.0cm, width=1\textwidth]{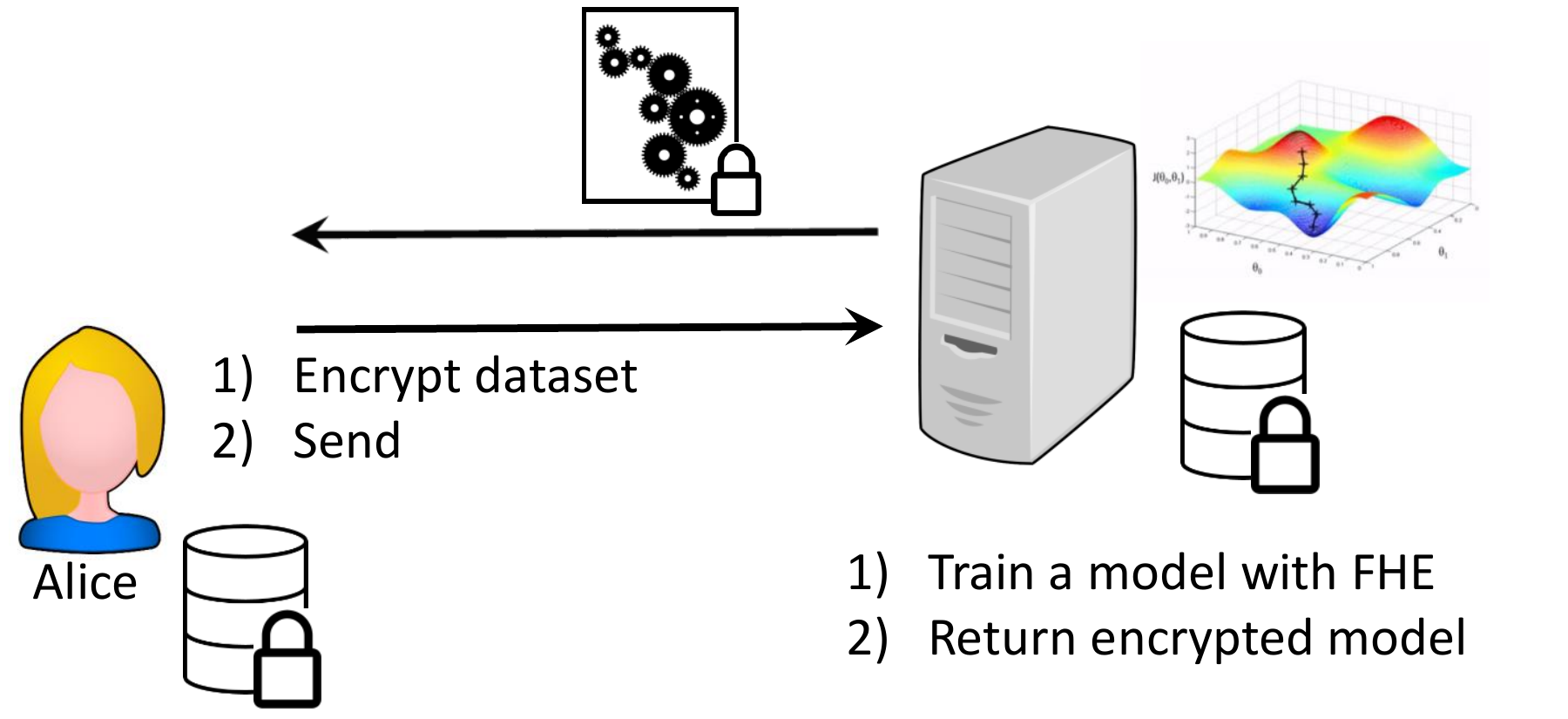}
        \caption{Secure training as a service.}
        \label{fig:training:as:a:service}
    \end{subfigure}
    \caption{Main tasks in text classification.}
    \label{fig:secure:spam:detection}
\end{figure*}

To give motivational use-cases for \privft{}, consider the inference and training as a service shown in Figure~\ref{fig:secure:spam:detection}. In the fully encrypted e-mail service shown in Figure~\ref{fig:inference:as:a:service}, Alice composes an e-mail and encrypts it using Bob's public key. The encrypted e-mail is sent to the mail server, which can still run a spam detection algorithm homomorphically with FHE. The encrypted result of spam detection and the still encrypted e-mail are forwarded to Bob, who can decrypt and decide whether to decrypt and open the e-mail or discard it based on the spam detection result. In Figure~\ref{fig:training:as:a:service}, Alice has a private dataset and wants to train a model on the cloud. She encrypts her dataset and sends to the cloud. The cloud runs the training algorithm and generates an encrypted model that is communicated back to Alice. Alice can decrypt the model and use it for local inference. Similar application domains can benefit from \privft{} use-cases such as targeted advertising, personalized marketing, recommendation systems, delegated digital watermarking/signature, etc.

To this end, we propose a secure system that is able to perform the two main tasks of text classification: 1) evaluating an existing model on encrypted inputs and 2) training a model using encrypted dataset. We employ a new FHE scheme, known as the Cheon-Kim-Kim-Song (CKKS) scheme~\cite{cheon2017homomorphic}, which has been proposed to deal naturally with floating-point numbers that require intricate treatment in conventional FHE schemes. To obtain practical performance, we leverage the computing power of GPUs. We implement a Residual Number System (RNS) variant~\cite{cheon2018full} of the CKKS scheme on GPUs to parallelize the core operations of the scheme. Moreover, we show how to extract extra parallelism at the application level using the packing methods provided by the CKKS scheme and task decomposition and scheduling to multi-GPU systems. To evaluate the performance of our system, we conduct several experiments to evaluate different aspects of the system. First, we compare the performance of our GPU implementation of the CKKS scheme with an existing CPU-based implementation provided by the Microsoft SEAL library~\cite{SEAL}. Secondly, we evaluate the accuracy of our system on 6 publicly available datasets and confirm that zero accuracy loss is noticed compared with the plaintext version of each task. Thirdly, we conduct a number of experiments to measure the runtime of each task in our system. Using GPU, the inference task can achieve a significant speedup for practical real-time applications. The runtime of the inference task is tested on 6 datasets ranges between 0.23 seconds to 0.66 seconds. As the training task is more computationally intensive, it is tested on 1 dataset recording 5.04 days of computation time on NVIDIA DGX-1 multi-GPU platform with 8 V100 cards.

Our main contributions can be summarized as follows:
\begin{itemize}
    \item To the best of our knowledge, we provide the first GPU implementation of an RNS variant of the CKKS \textit{levelled} FHE scheme.
    \item We demonstrate how homomorphic inference can be performed on encrypted data for text classification.
    \item We demonstrate how to train an effective model using an encrypted dataset with FHE.
    \item We provide extensive benchmarking experiments to evaluate the performance of our system.
\end{itemize}

\subsection{Organization of the paper}
The rest of the paper is organized as follows. We review the state of the art related to our work in Section~\ref{sec:related:work}. Section~\ref{preliminaries} reviews the basic terminology and concepts the paper builds on. A description of how \privft{} performs both inference and training on encrypted data is presented in Section~\ref{PrivFT}. Section~\ref{Implementation} provides the implementation details of the CKKS scheme and \privft{}. We present our experimental results in Section~\ref{Performance:Evaluation}. Finally, Section~\ref{sec:conclusion} summarizes the work and provides directions for future work.

\section{Related Work}\label{sec:related:work}
The interest in privacy-preserving ML has gained a substantial momentum in the last decade due to: (1) the inclusion of ML in high-stakes domains such as healthcare, autonomous vehicles and finance~\cite{dua2014machine,goodall2014machine,dunis2016artificial}, (2) the increased awareness of the misuse of private data~\cite{isaak2018user}, and (3) the emergence of privacy-preserving cloud computing technologies that enable computing on encrypted data such as FHE and Multi-Party Computation (MPC)~\cite{yao1982protocols}. While MPC solutions require less computational overhead, they require high interaction between the users and cloud, hence they are bandwidth-bound. On the other hand, FHE offers non-interactive solutions at the cost of high computational requirements. Clients need only to be online to provide inputs and receive outputs to/from the cloud. These technologies have been used to construct several privacy-preserving application~\cite{du2001secure,ARXIV:AslEspHol15,MSFT:DGL+16,archer2018keys,juvekar2018gazelle,badawi2018alexnet,chou2018faster}, we briefly review here two studies that addressed private text classification with FHE and MPC.

In~\cite{costantino2017privacy}, the authors provide a secure text classifier using FHE in which a data analyzer (say Bob) learns the frequency of occurrences for a set of words he has, in the data provider (Alice)'s text, without learning the text. This is enabled by relying on a third semi-honest\footnote{A semi-honest party is eager to learn private data but does not deviate from the specification of the underlying protocol.} party they called HE engine (HEng). In this system, Alice sends her private text encrypted using Bob's private key to HEng. Bob sends his private list of words encrypted under his own private key to HEng. HEng computes the word frequencies using FHE and returns the encrypted result to Bob who decrypts it and uses it as input to a text classifier. Note that Bob can learn all the words (out of order) in Alice's text if he provides all the words in the English dictionary to HEng. Moreover, the classifier runs on plaintext data in contrast to our solution which runs entirely on an encrypted encoding of the text. In terms of performance, their system requires 19 minutes to analyze a single tweet and 78 minutes to analyze an email at high accuracy ($>$94\%).

Another related work has been proposed recently in ~\cite{de2019privacy} using MPC. The authors propose a system to classify text messages using two algorithms: Logistic Regression (LR) and ensemble trees. The main building block of this solution is a protocol for doing secure equality test which is used to perform secure comparisons and feature extraction. Unlike the previous solution, the classifiers in this solution run on secret-shared data. The authors tested their solution on a dataset consisting of 10,000 tweets~\cite{i2019multilingual}, with 60\% being tagged as hate speech. In terms of performance, this system achieved better latency results 28.3 seconds but at poor accuracy 74.4\%. As will be shown in Section~\ref{Performance:Evaluation}, our results show that we can classify unstructured texts (from different standard datasets) in less than a second at much higher accuracy 91.49\% - 98.80\%.

\section{Preliminaries}\label{preliminaries}
In this section, we review the basic mathematical notions our paper builds on. We start by describing the notations used, followed by a brief introduction of FHE and text classification as an NLP problem.

\subsection{Notations}
We use capital letters to refer to sets and small letters for elements of a set. The sets $\mathbb{Z}, \mathbb{R},$ and $\mathbb{C}$ denote the integers, reals and complex numbers, respectively. We use capital bold letters for matrices and bold small letters for vectors. The symbols $\lceil \cdot \rceil, \lfloor \cdot \rfloor$, and $\lfloor \cdot \rceil$ denote the round up, round down and round to nearest integer, respectively. The notation $|a|_{q}$ denotes the remainder of $a$ divided by $q$. We use $|A|$ to denote the size of set A. Finally, sampling $a$ from a set $\mathcal{S}$ is denoted as $a \xleftarrow{} \mathcal{S}$.

\subsection{Fully Homomorphic Encryption}\label{sec:FHE}
FHE schemes are cryptographic constructions that provide the ability to compute on encrypted data without decryption~\cite{STOC:Gentry09}. Unlike classic encryption schemes, FHE maps the input clear text data (or plaintexts $\mathcal{P}$) to encrypted data (or ciphertexts $\mathcal{C}$) such that the algebraic structure between $\mathcal{P}$ and $\mathcal{C}$ is preserved over addition and multiplication. Let $a, b \in \mathcal{P}$ and $Enc$ denotes the encryption operation, then $Enc(a) \oplus Enc(b) = Enc(a+b)$ and $Enc(a) \odot Enc(b) = Enc(a \cdot b)$, where $\oplus$ and $\odot$ are homomorphic addition and multiplication, respectively, and where equality is achieved after decryption. This allows one to evaluate arbitrary computations (modeled as circuits) on encrypted data by only manipulating the ciphertexts. 

Modern FHE schemes conceal plaintext messages with noise that can be identified and removed with the secret key~\cite{brakerski2011fully}. As we compute on encrypted data, the noise magnitude increases at a certain rate (high rate for multiplication and low rate for addition). As long as the noise is below a certain threshold, that depends on the encryption parameters, decryption can filter out the noise and retrieve the plaintext message successfully. Although FHE schemes include a primitive (known as \textit{bootstrapping}) to refresh the noise ~\cite{STOC:Gentry09}, it is extremely computationally intensive. Instead, one can use a \textit{levelled} FHE scheme~\cite{brakerski2014leveled} that allows evaluating circuits of multiplicative depth\footnote{The maximum number of multiplications along any path in the evaluated function. In FHE, multiplication is more expensive than addition, hence it is crucial to optimize the number of multiplications.} below a certain threshold, which can be controlled by the system parameters. The literature includes various FHE schemes that vary in the underlying mathematical structures used, capabilities and performance. This work uses the (CKKS) \textit{levelled} FHE scheme~\cite{cheon2017homomorphic}, which was proposed specifically to deal with floating-point numbers.

Although FHE is known for being slow~\cite{naehrig2011can}, a number of major advancements have improved its performance dramatically such as (1) packing methods~\cite{DCC:SmaVer14}, which allow one to pack a vector of plaintext items in one ciphertext enabling vectorized homomorphic operations without extra cost, (2) fast modular arithmetic that replaces slow multi-precision operations with embarrassingly parallel native operations~\cite{bajard2016full,EPRINT:HalPolSho18,cheon2018full}, and (3) hardware acceleration via GPUs~\cite{TCHES:BVMA18,8657794} which can provide 1 to 2 orders of magnitude against CPU implementations.

\subsection{Text Classification}
Text Classification is a task in NLP with numerous applications such as Sentiment Analysis, Spam Detection, Topic Classification and Document Classification. Much of the recent NLP research has focused on transfer learning techniques such as pre-training word embeddings~\cite{pennington2014glove,mikolov2013efficient}, or pre-training language models on larger datasets and fine-tuning them for task-specific learning~\cite{radford2017learning,peters2018deep,devlin2018bert,howard2018universal}. \textsf{SentimentUnit}~\cite{radford2017learning} employs a single layer multiplicative Long Short Term Memory (LSTM) while \textsf{ELMO}~\cite{peters2018deep} adopts a base architecture which contains multiple layers of Bidirectional LSTM. \textsf{BERT}~\cite{devlin2018bert} on the other hand uses the Transformer architecture~\cite{vaswani2017attention} which consists of stacked attention layers. In \textsf{ULMFit}, the AWD-LSTM~\cite{merity2017regularizing} is used for pre-training and at fine-tuning time is able to achieve impressive results across a variety of text classification tasks. In \cite{sun2019fine}, the authors used the Google's $\text{BERT}_{\text{Large}}$ model \cite{devlin2018bert} and further experimented with several fine-tuning and task-specific pre-training techniques to boost the accuracy levels. Finally, \cite{yang2019xlnet} proposes the XLNet model to better capture the bidirectional dependency of words that's inherent in natural language texts.  

While these approaches achieved remarkable results across many NLP tasks, they are prohibitively expensive for FHE adaptation. As mentioned above, FHE schemes typically introduce a noise term in the encrypted ciphertext which grows with every addition or multiplication operation. Depending on the encryption parameters, the decryption of such ciphertext only yields the correct result when this noise term is small enough. As a result, both the recurrence functions (e.g. of LSTM) as used in \textsf{SentimentUnit, ELMO, ULMFit, XLNet} and the stacked Transformer attentions in \textsf{BERT, XLNet} result in a level of multiplicative depth that would corrupt the result of the computations once decrypted. 

\begin{figure}[!htpb]
    \centering
    \includegraphics[width=1\columnwidth,height=4.8cm]{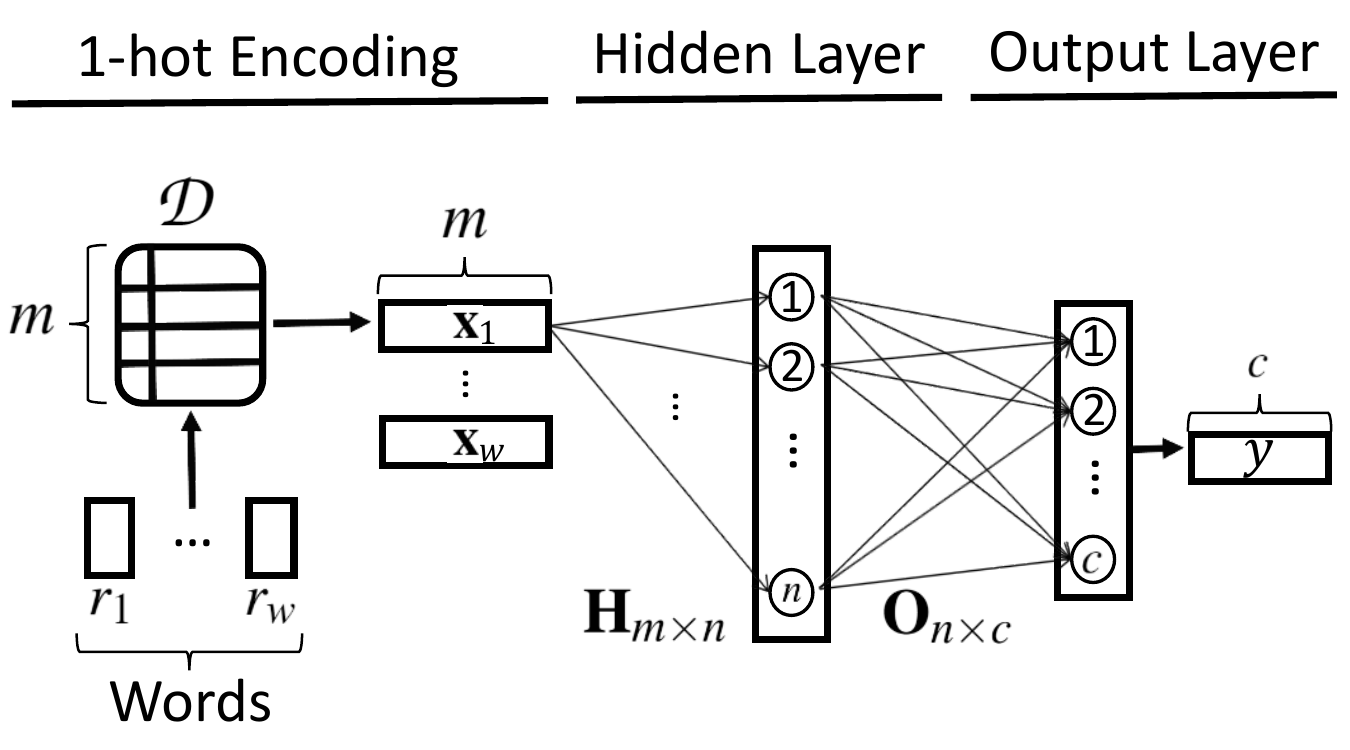}
    \caption{Network architecture of vanilla \fasttext{}.}
    \label{fig:fasttext:vanilla}
\end{figure}

Motivated by this limitation, we choose \fasttext{}~\cite{joulin2016bag} for our work as it is a shallow network consisting of only two layers: an embedding (hidden) layer and an output fully connected layer as shown in Figure~\ref{fig:fasttext:vanilla}. The input to the model is a vector of words $r_1, \ldots, r_w$. Firstly, each word $r_i$ is encoded via a 1-hot encoding scheme with the aid of a dictionary $\mathcal{D}$ of length $m$, where $m$ is the total number of words in the dictionary. Via a lookup operation, word $r_i$ is replaced by a vector $\mathbf{x}$ of length $m$ such that $\mathbf{x}_j$ is set to 1, where $j$ is the word index in the dictionary. The remaining components of $\mathbf{x}$, i.e., $\mathbf{x}_i$, where $i \neq j$,  are set to zero. The embedding layer uses a hidden lookup matrix $\mathbf{H}$ of size $m \times n$, where $n$ is the dimension of the floating-point embedding vector encoding the input word. Each row in $\mathbf{H}$ corresponds to a word in the dictionary. The lookup vectors are averaged over all $w$ words to produce a single vector $\mathbf{h} = \frac{1}{w} \sum_i (\mathbf{x}_i) \cdot \mathbf{H} \in \mathbb{R}^n$. The output layer uses a matrix $\mathbf{O}$ of size $n \times c$ to map the result into the output space containing $c$ classes. For training, a \textsf{softmax} function is used to compute the loss function. At inference time, however, one only needs to compute the unnormalized scores $\mathbf{s} = \mathbf{h} \cdot \mathbf{O}$. Thus, the multiplicative depth in a single inference pass is only 3 (2 vector-matrix multiplications and average computation) which is an attractive level for FHE implementation.

It should be noted that, firstly, word embedding refers to a fundamental procedure in NLP where words are converted into floating-point vector representations in a manner that captures the context of a word in text, both semantic and syntactic similarity, and relation with other words~\cite{mikolov2013distributed}. More specifically, the similarity of objects is reflected in their similarity in the embedding space, usually measured by distance. Secondly, \fasttext{} has two embedding models: 1) skip-gram and 2) continuous bag of words, where only the latter is used in our work.


\subsection{The CKKS \textit{levelled} FHE Scheme}
In the following paragraphs, we briefly describe an RNS version of the \textit{levelled} CKKS scheme~\cite{cheon2017homomorphic}, which is used in our implementation. 

Let $N$ be a power of 2 integer and $R=\Z[X]/(X^N+1)$ be the ring of polynomials modulo $(X^N+1)$, having degree at most $N-1$ and integer coefficients. For some positive integer $q$, let $R_q=\Z_q[X]/(X^N+1)$ consists of the same polynomials but having coefficients modulo $q$. Let $q_L > q_{L-1} > \cdots > q_1$ be $L$ positive integers such that $q_l = \prod_{i=1}^{l} p_i$, where $p_i$'s are small prime integers. We refer to a prime that fits in a single machine word as a small prime, i.e., up to 32-bit or 64-bit primes on current commodity machines. In CKKS, $L$ is the maximum multiplicative depth supported (hence ``\textit{levelled}''), and at any level $l=1 \ldots L$, all arithmetic operations are performed in $R_{q_l} = \Z_{q_l}[X] / \langle X^N+1 \rangle$. When the input data is first encrypted, the ciphertext polynomials are elements of $R_{q_L}$ (at level $L$), but may later need to be rescaled into $R_{q_{L-1}}$. Conceptually, one can think of CKKS as having $L$ levels where computation begins at the highest level $L$ and moves down to a lower level as needed (more formal discussion to follow). Denote DFT and IDFT as the Discrete Fourier Transform and its inverse, respectively. Given input data represented as real or complex numbers, we use a modified version of CKKS~\cite{cheon2018full} for encryption (\textsf{ENC}) and decryption (\textsf{DEC}) as follows:
\begin{itemize}
    \item \textsf{SETUP}: given a desired security level\footnote{The security level ($\lambda$) corresponds to the computational effort required to break the scheme with probability 1 using $2^{\lambda}$ elementary operations.} $\lambda$, and maximum computation levels $L$, initialize CKKS by setting $N$, two uniform random distributions: $\mathcal{X}_{key}$ over $R_2$ and $\mathcal{X}_{q_{L}}$ over $R_{q_{L}}$, and a zero-centered discrete Gaussian distribution $\mathcal{X}_{err}$ with standard deviation $\sigma$ over $R_{q_{L}}$.
    \item \textsf{KEYGEN}: generate: (1) secret key $s \leftarrow \mathcal{X}_{key} \in R_{q_L}$, and (2) public key $(a,b) \in R_{q_L}^2$, where $a \leftarrow \mathcal{X}_{q_{L}}$ and $b = -as + e$ with $e \leftarrow \mathcal{X}_{err}$.
    \item \textsf{ENCODE}($v, \rho$): given a vector of complex numbers $\mathbf{v} \in \C^{N/2}$ and precision $\rho$, return a polynomial $\mu = \lfloor \textsf{IDFT}(2^\rho v) \rceil \in R$.
    \item \textsf{ENC}$(\mu)$: given a plaintext message ($\mu$), sample $u \leftarrow \mathcal{X}_{q_{L}}$ and $e_0, e_1 \leftarrow \mathcal{X}_{err}$. Return ciphertext $ct = (c_0, c_1) = ( av + \mu + e_0, bv + e_1 ) \in R_{q_{L}}^2$.
    \item \textsf{DEC}$(ct)$: given a ciphertext $ct \in R_{q_{l}}^2$, return $\mu = c_0 + sc_1 \in R_{q_l}$.
    \item \textsf{DECODE}($\mu, \rho$): given $\mu \in R$ and precision $\rho$, return $v = \textsf{DFT}(\mu/2^\rho) \in \C^{N/2}$.
\end{itemize}

To enable computations in the ciphertext space, the following homomorphic operations are given: 
\begin{itemize}
    \item \textsf{HADD}$(ct_0, ct_1)$: homomorphic addition takes two ciphertexts (at the same level $l$) and returns $ct^{+} = ct_0 + ct_1 \in R_{q_l}^2$.
    \item \textsf{HMUL}$(ct_0 = (c_{00}, c_{01}), ct_1 = (c_{10}, c_{11}))$: homomorphic multiplication takes two ciphertexts (at the same level $l$) and returns $ct^{\times} = (c_{00}c_{10}, c_{00}c_{11} + c_{01}c_{10} , c_{01}c_{11}) \in R_{q_l}^3$. Note that a procedure known as relinearization~\cite{cheon2017homomorphic} can be used to reduce $ct^{\times}$ back to two elements $\in R_{q_l}^2$.
    \item \textsf{HADDPLAIN}$(ct, pt)$: homomorphic addition of a ciphertext $ct = (c_0, c_1) \in R_{q_{l}}^2$ and plaintext $pt \in R$ returns ciphertext $ct^{+} = (c_0+pt, c_1) \in R_{q_{l}}^2$.
    \item \textsf{HMULPLAIN}$(ct, pt)$: homomorphic multiplication of a ciphertext $ct = (c_0, c_1) \in R_{q_{l}}^2$ and plaintext $pt \in R$ returns ciphertext $ct^{\times} = (c_0 \cdot pt, c_1 \cdot pt) \in R_{q_{l}}^2$.
\end{itemize}

CKKS mimics fixed-point arithmetic for approximate computing on encrypted numbers. Input real numbers are scaled with a fixed-precision factor and rounded to the nearest integer (quantization). For instance, the value 3.14159 can be represented as 3142 with a scale factor of $1/1000$. To maintain a fixed precision of the intermediate values, CKKS offers an efficient \textit{rescaling} procedure (\textsf{RESCALE}) to remove the least significant bits of intermediate results. For instance, after multiplying two messages $m_1$ and $m_2$ each scaled with factor $\rho$, \textsf{RESCALE} produces a rounded version of the product $\rho \cdot \lfloor 1/\rho \cdot m_1 m_2 \rceil$ instead of $\rho^2 \cdot m_1 m_2$:

\begin{itemize}
    \item \textsf{RESCALE}$(ct, l')$: given ciphertext at level $l$ and $l' = l - 1$, return $ct' = \lfloor q_{l'}/q_l \cdot ct \rceil \in R_{q_{l'}}$.
\end{itemize}

As mentioned in section \ref{sec:FHE}, one can drastically improve FHE performance via packing methods. In CKKS, a vector of up to $N/2$ complex numbers can be encoded in a single plaintext element. This allows one to perform Single-Instruction Multiple-Data (SIMD) homomorphic operations on packed ciphertexts for free. Packing can be viewed as if the ciphertext has independent slots, each concealing one data item. Adding or multiplying two packed ciphertexts results in a packed ciphertext that contains the component-wise summation or product of the input ciphertexts. To facilitate later discussion on packed ciphertexts arithmetic, we will need the following utility function:

\begin{itemize}
    \item \textsf{ROTATE}$(ct, \pi \in \Z)$: the packing slots inside a ciphertext can be rotated ($\pi$ slots to the left or right depending on the sign of $\pi$) by computing $ct' = (c_0(X^\pi), c_1(X^\pi))$.\footnote{The rotated ciphertext $ct'$ can only be decrypted using $s' = s(X^\pi)$. To make $ct'$ decryptable under the original secret key $s$, a procedure known as key switching~\cite{cheon2017homomorphic} is used with an associated rotation/switching key that can be generated at system initialization.}
\end{itemize}

The reader is referred to the referenced papers for proofs on the correctness and security of the scheme.

\section{\privft{}}~\label{PrivFT}
In this section, we discuss the main two tasks in \privft{}: homomorphic inference and training with FHE. We first describe briefly how they are performed in \fasttext{}. Next, we present our adaptations to the procedures so they can be evaluated homomorphically with FHE.

\subsection{Homomorphic Inference with FHE}

We first define a dictionary $\mathcal{D}$ of length $m$ as a set of key-value tuple. The keys are integers running from $0$ to $m-1$. Each key is associated with a single value that is a word from a specific language. We define $\mathcal{D}(r)$ as the lookup operation in the dictionary returning the index of word $r$. The inference task of \privft{} is shown in Algorithm~\ref{alg:fasttext:forward:propagation}.
\begin{algorithm}[!ht]
         \caption{\textsf{\fasttext Inference}} \label{alg:fasttext:forward:propagation}
         \begin{algorithmic}[1]
             \Statex{\textbf{Input}: a dictionary $\mathcal{D}$ of length $m$ words, set of words $r_1, \ldots, r_w$, embedding matrix $\mathbf{H}$ of size $m \times n$, where $n$ is the dimension of the embedding space, and output matrix $\mathbf{O}$ of size $n \times c$, where $c$ is the number of classes.} 
             \Statex{\textbf{Output}: a floating-point vector $\mathbf{y}$ of length $c$ that corresponds to the probabilities $p(c_i|r_1,\ldots, r_w)$}
             
             \Statex{\textbf{Step 1: 1-hot Encoding}}
             \For{$i = 1$ to $w$ }
                \State{$\mathbf{x}_i = \{0,\ldots,0\}$}
                \State{$idx = \mathcal{D}(\mathbf{x}_i)$}
                \State{$\mathbf{x}_i(idx) = 1$}
             \EndFor
             \Statex{\textbf{Step 2: Compute embedded representation}}
             \State{$\mathbf{h} = \frac{1}{w} \sum_{i=1}^{w} \mathbf{x}_i \cdot \mathbf{H}$}
             
             \Statex{\textbf{Step 3: Compute class scores}}
             \State{$\mathbf{s} = \mathbf{h} \cdot \mathbf{O}$}
             
             \Statex{\textbf{Step 4: \textsf{softmax}}}
             \For{$i = 1$ to $c$ }
                \State{$\mathbf{y}_i = \frac{exp(\mathbf{s}_i)}{\sum_{j=1}^{c}exp(\mathbf{s}_j)}$}
             \EndFor
             \State \textbf{return} $\mathbf{y}$
         \end{algorithmic}
\end{algorithm}

Ideally, \fasttext{} homomorphic inference should be executed as follows: 1) client encrypts the words of her input text and sends them to the server, 2) the server would evaluate steps 2 to 4 in Algorithm~\ref{alg:fasttext:forward:propagation} homomorphically and sends the result back to the client. Unfortunately, with the current state of FHE, this scenario is not practical due to the limited arithmetic set currently provided by most FHE schemes. For instance, homomorphic lookup in step 3 is an expensive operation that requires large multiplicative depth depending on the size of the dictionary~\cite{chen2017fast}. Another challenge arising in Algorithm~\ref{alg:fasttext:forward:propagation} is the evaluation of \textsf{softmax} in FHE. This is even more complex compared to the lookup operation as it includes evaluation of transcendental functions and division which require large circuits depth to evaluate. Due to the previously mentioned reasons, we adapt the inference task in \fasttext{} to obtain practical performance.

Firstly, to adapt the initial lookup in \fasttext{} to be more FHE-friendly, we require the client to encode her text $r_1,\ldots,r_w$ into a 1-hot vector ($\mathbf{v}$) of length $m$, where $\mathbf{v}_i = 1$ at all indices returned by $\mathcal{D}(\mathbf{r}_i),~\forall i \in \{1,\ldots,w\}$. More concretely, $\mathbf{v} = \sum_{1}^{w} \mathbf{x}_i$. Note that we abuse the definition of 1-hot encoding here as $\mathbf{v}$ might contain more than one non-zero values. Besides, a repeated word at dictionary index $i$ results in values greater than 1 in $\mathbf{v}_i$. The client encrypts $\mathbf{v}$ and sends it along with the number of words $w$ (in plaintext) to the server. By doing so, the server can compute the embedded representation $\mathbf{h}$ (Step 2) by a vector-matrix multiplication ($\mathbf{v} \cdot \mathbf{H}$) and a plaintext multiplication (\textsf{HMULPLAIN}) of the factor $1/w$. We assume here that the embedding vectors in $\mathbf{H}$ correspond to the words in the same order as indexed in the dictionary. This is considered a weak assumption since the dictionary can be fixed before the embedding representation is trained.

\begin{figure*}[!ht]
    \centering
    \includegraphics[width=1.00\textwidth, height=7.5cm]{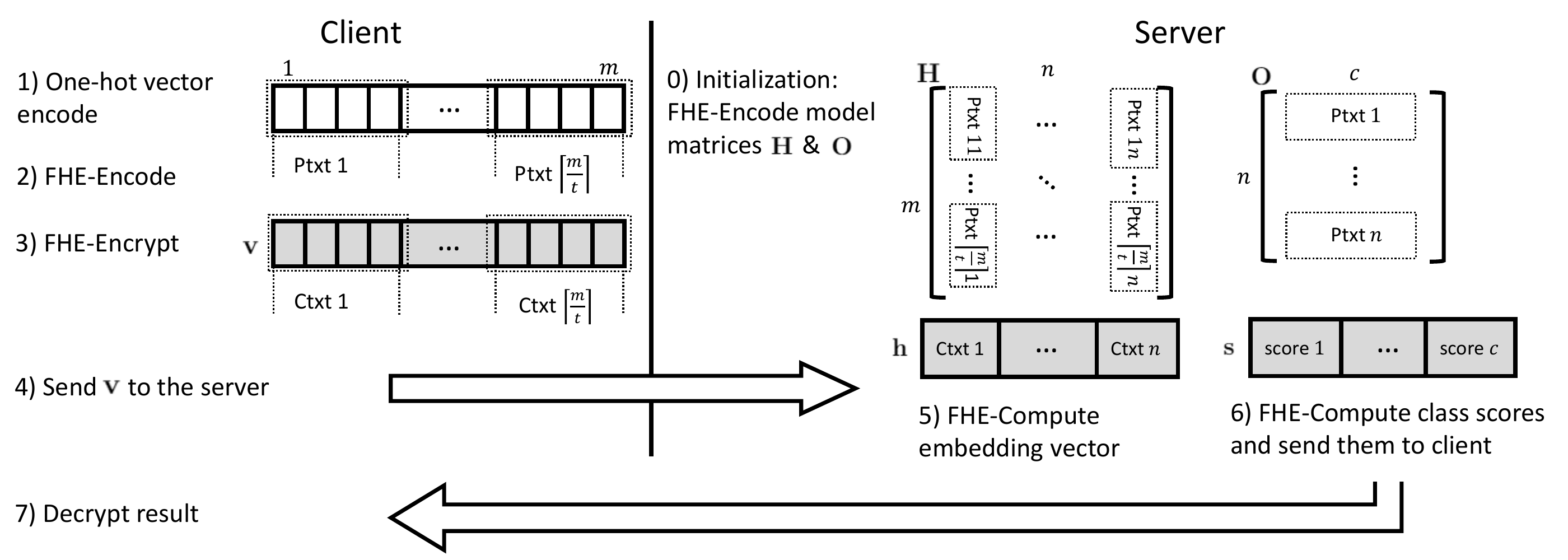}
    \caption{\privft{} prediction task. Ptxt and Ctxt refer to plaintext and ciphertext, respectively. $t$ denotes the number of slots in plaintext or ciphertext. Shaded boxes are ciphertexts while clear ones are plaintexts.}
    \label{fig:fasttext:he:prediction}
\end{figure*}

An important optimization we introduce here is related to matrix packing. Packing proves to be useful in FHE to reduce both the number of homomorphic operations due to SIMD evaluation and the number of ciphertexts. We use packing at both the client and server sides. At the client, $\mathbf{v}$ is stored in $\lceil m/t \rceil$ plaintext objects, where $t$ is the number of slots in plaintext/ciphertext\footnote{The maximum number of slots in plaintext/ciphertext is $N/2$, where $N$ is the ring dimension.}. At the server, since $m$ is usually large, we pack the hidden matrix $\mathbf{H}$ vertically to perform the dot-product efficiently. Specifically, the server encodes the weights matrices of the hidden and output layers vertically and horizontally, respectively as shown in Figure~\ref{fig:fasttext:he:prediction}. To encode $\mathbf{H}$, we use $n \times \lceil m/t \rceil$ plaintexts. For the output matrix $\mathbf{O}$, we assume that the number of classes $c$ is less than $t$ and pack the weights horizontally requiring only $n$ plaintexts. If $c > t$, multiple plaintexts can be used. Note that our assumption is reasonable as the number of classes in common text classification tasks is usually small ($\ll N/2$).

To compute ($\mathbf{v} \cdot \mathbf{H}$), the dot-product is computed via element-wise ciphertext-plaintext homomorphic multiplication $\mathbf{h}_j' = $ \textsf{HMULPLAIN}$(\mathbf{v}_i, \text{Ptxt}_{j,i})$, followed by $\lceil \frac{m}{t} \rceil$ \textsf{HADD}$(\mathbf{h}_j', \mathbf{h}_{j+1}')$ to generate a single ciphertext $ct$. The elements in the slots of $ct$ can be summed to generate $\mathbf{h}_j$ using the \textsf{TotalSum} procedure~\cite{halevi2014algorithms} in $\mathcal{O}(\log{N})$ time complexity, as shown in Algorithm~\ref{alg:tot:sum}. Note that component $\mathbf{h}_i$ contains one ciphertext that encrypts single value which is the dot product between $\mathbf{v}$ and the $i$-th column in $\mathbf{H}$. 

To compute the class scores $\mathbf{s}$, a slightly different approach is used to multiply $\mathbf{h}$ with the output matrix $\mathbf{O}$. Since $\mathbf{h}$ is non-packed, we multiply each component $\mathbf{h}_i$ by the rows of $\mathbf{O}$ to generate $n$ packed ciphertexts. This step requires $n$ \textsf{HMULPLAIN} operations. The resultant ciphertexts are summed to generate a packed ciphertext encrypting $\mathbf{s}$, which is communicated back to the client who can decrypt and find the best class by evaluating the \textsf{softmax} function in plaintext.

We remark that as the server performs FHE-friendly operations (addition and multiplication), \privft{} does not suffer from any accuracy loss in the inference task.
\begin{algorithm}[!ht]
         \caption{\textsf{TotalSum}} \label{alg:tot:sum}
         \begin{algorithmic}[1]
             \Statex{\textbf{Input}: ciphertext $ct$ encrypting vector $\mathbf{v}$ and number of slots $t$}
             \Statex{\textbf{Output}: a ciphertext encrypting the total sum of elements in $v$, duplicated in slots}
             \For{$i = 1$ to $\log_2{N}$ }
                \State{$t =$ \textsf{ROTATE}$(ct, 2^i)$)}
                \State{$ct = $ \textsf{HADD}($ct$, $t$)}
             \EndFor
             \State \textbf{return} $ct$
         \end{algorithmic}
 \end{algorithm}

\subsection{Homomorphic Training with FHE}

In order to train the network with FHE, a number of challenges have to be addressed. First, the maximum multiplicative depth of the training circuit should be minimized in order to avoid \textit{bootstrapping}. We note that training an effective model on encrypted data with FHE has traditionally not been possible, either on CPU or GPU, due to the following reasons: (1) the multiplicative depth is roughly doubled (compared to inference time) for each epoch due to the backward pass, (2) it is further increased by the (typically) many training epochs, (3) the network architecture may already require a large depth for a single pass. Therefore, our third contribution is mainly how to reduce the multiplicative depth for the training task, while still training an \textit{effective} model. Before we delve into our adaptations, we briefly describe the training procedure in \fasttext{}.

The training procedure is used to find the weights of the hidden and output layers $\mathbf{H}$ and $\mathbf{O}$. The pseudocode of the basic training procedure, known as Gradient Descent (GD), is shown in Algorithm~\ref{alg:training}. \fasttext{} uses Stochastic GD (SGD) to update the model after every prediction of each data-point in the dataset. This style of training is not suitable in FHE as it requires an extremely large multiplicative depth. Updating the model can be simplified as $\text{model}_{\tau+1} = \text{model}_{\tau} - \eta \cdot \text{gradient}$, where $\eta$ is the learning rate. The gradient is computed as an evaluation of the loss function on the prediction (inference) of an input example. This requires a certain multiplicative depth. Subsequent predictions use the updated model in the inference and deepen further the multiplicative depth. 

Another training style, known as minibatch SGD, divides the training dataset into batches, accumulate the gradients in each batch and perform a single update to the model for each batch. This procedure results in a much lower homomorphic multiplicative depth compared to SGD. Thus, we adapt \fasttext{} to use minibatch SGD instead of true SGD and design \privft{} to use the same approach.

Here, we observe that the size of the minibatch is inversely proportional to the multiplicative depth required. Thus, we use gradient descent with large minibatch size and fewer epochs to reduce the number of weight updates and accordingly the circuit multiplicative depth for FHE evaluation. Also, we use a small number of epochs to train the model. 
\begin{algorithm}[!ht]
         \caption{\textsf{GDTraining}} \label{alg:training}
         \begin{algorithmic}[1]
         \Statex{\textbf{Input}: training dataset (train\_data)}
         \Statex{\textbf{Output}: a trained model (model)}
         \State{model = initialize()} \Comment{Often randomly}
             \For{$i = 1$ to $\text{num\_epochs}$ }
                \State{X = shuffle(train\_data)} \Comment{predictors}
                \State{y = ground\_truth(X)} \Comment{dependent variable}
                \State{predictions = predict(X, model)} \Comment{inference}
                \State{error = calculate\_error(y, predictions)}
                \State{model = update\_model(model, error)}
             \EndFor
             \State \textbf{return} model
         \end{algorithmic}
 \end{algorithm}

The second challenge is how to evaluate the loss function with FHE. In \fasttext{} the authors use \textsf{softmax} as a loss function. Since FHE provides only addition and multiplication, evaluating \textsf{softmax} as a circuit can be very expensive. Instead, we use a shallow approximation polynomial that can be evaluated cheaply with FHE. Using the Minimax approximation algorithm we found that the polynomial $(\frac{1}{8}X^2 + \frac{1}{2}X + \frac{1}{4})$ can be used as a good approximation to \textsf{softmax}. In terms of accuracy, the polynomial approximation showed no noticeable accuracy loss. We note for some problems, it has been showed that polynomial approximation may generate accurate results (see \cite{MSFT:DGL+16, badawi2018alexnet}.

Combined with the simplicity of the \fasttext{} model, the approach we proposed is theoretically sound and is to our knowledge the first to train a strong NLP model entirely on encrypted data. However, in practice, large system computational requirements are needed. For instance, memory requirements of the training task on encrypted dataset could not be met by our GPU system as the system memory was not sufficient. To overcome this problem, we had to use tiling techniques to decompose the computation into smaller tasks and ensure that the required data migrates along the path from CPU memory to GPU memory and vice versa.

\section{Implementation of \privft{}}\label{Implementation}
In this section, we provide the implementation details of \privft{}. We start by describing the GPU implementation of the CKKS scheme. Next, we describe the details of our implementation of both inference and training tasks.

\subsection{GPU implementation of CKKS}

We implement an RNS variant~\cite{cheon2018full} of the CKKS \textit{levelled} FHE scheme~\cite{cheonhomomorphic} using CUDA 10. The usage of RNS allows implementing the scheme using native 32- and 64-bit operations with high parallelism instead of slow, serial multi-precision operations. Core polynomial arithmetic, RNS and Discrete Galois Transform (DGT) tools were obtained from the \ahe{} GPU library~\cite{TCHES:BVMA18,8657794} for high parallelism and improved performance. Below is a brief description of our implementation of the new tools required by CKKS.

We instantiate CKKS again on the cyclotomic polynomial ring $R$ and use ciphertext moduli $q_L > q_{L-1} > \cdots > q_1$. At any level $l$, arithmetic is performed in $R_{q_l} = \Z_{q_{l}}[X] / \langle X^N+1 \rangle$ modulo both $q_l$ and $\langle X^N + 1 \rangle$. 
Our implementation of the CKKS utilizes the core polynomial arithmetic provided by \ahe{} which implements the Fan-Vercauteren (FV) FHE scheme~\cite{EPRINT:FanVer12}. The major changes, however, are the encoding, decoding and rescaling procedures. Since \textsf{ENCODE} and \textsf{DECODE} are not on the critical path of homomorphic computation (they are invoked by the client before providing inputs and after receiving results to/from the cloud), we implement them in CPU. On the other hand, \textsf{RESCALE}, which is performed by the cloud during homomorphic computation, is critical to the performance. Implementing \textsf{RESCALE} in RNS is both complex and expensive as it requires comparison, which is hardly compatible with RNS. Instead, we implement it by computing the floor function using RNS division by a subset of the RNS moduli~\cite{omondi2007residue,EPRINT:FanVer12} as described in Equation~\eqref{eq:rns:floor}. The error generated from this approximation is added to the inherent noise included in FHE and found to be negligible by experimental results. Our implementation of \textsf{RESCALE} can be used to divide by the last prime modulus in the prime chain $p_l$. The prime chain is ordered such that $p_i < p_{i-1} $ to provide a simpler and more efficient scaling procedure, as shown in Algorithm~\ref{alg:rescale:rns}. Note that our ordered prime chain alleviates the need for expensive RNS base extension~\cite{omondi2007residue}.
\begin{equation}\label{eq:rns:floor}
    \biggr\lfloor \frac{q_l'}{q_l} \cdot c \biggr\rfloor = \biggr\lfloor \frac{1}{p_l} \cdot c \biggr\rfloor = \frac{(c - |c|_{p_l})}{p_l} \Mod{q_l'}
\end{equation}

\begin{algorithm}[!ht]
         \caption{RNS \textsf{RESCALE} by a single RNS modulus.} \label{alg:rescale:rns}
         \begin{algorithmic}[1]
             \Statex{Let $\mathcal{B}$ denote the underlying RNS base of $q_l = \{p_1, \cdots, p_l\}$. }
             \Statex{\textbf{Input}: $ct = (c_0, c_1) \in R_{q_l}^2$ in RNS representation and $l' = l - 1$}
             \Statex{\textbf{Output}: $ct' = (c_0', c_1') = \lfloor q_l'/q_l \cdot ct \rceil$}
             \For{$i = 1$ to $2$ }\Comment{For each polynomial in $ct$}
                \For{$j = 1$ to $N$ }\Comment{For each coefficient}
                    \For{$k = 1$ to $l'$ }
                        \State{$c'_i[j] = ({c_i[j]}_{p_k} - {c_i[j]}_{p_l}) \times p_l^{-1}\Mod{p_k} $}\Comment{Since $p_l$ is less than $p_k$, ${c_i[j]}_{p_l}$ is used without RNS base extension.  }
                    \EndFor
                \EndFor
             \EndFor
             \State \textbf{return} $ct'$
         \end{algorithmic}
     \end{algorithm}

\subsection{\privft{} Implementation}

\privft{} is implemented in two libraries on two different hardware platforms. The first implementation uses Microsoft SEAL v3.3.0 and runs on CPU. The second implementation utilizes our GPU implementation of the CKKS scheme - described in the previous section and runs on NVIDIA-enabled GPUs. The CPU implementation of \privft{} is fairly straightforward as we only use additions, multiplications and rotations. On the other hand, the GPU implementation is slightly more complex as the GPU memory is quite limited. We focus here on the GPU implementation and show how we tile and exploit parallelism in inference and training. We note that the parallelism techniques are used in both CPU and GPU implementations.

\subsubsection{Homomorphic Inference Implementation}
One can refer to Figure~\ref{fig:fasttext:he:prediction} to understand the implementation of homomorphic inference in ~\privft{}. An encrypted vector $\mathbf{v}$ is first received from the client. The server evaluates two vector-matrix multiplications. As the model is unencrypted, 3 \textsf{HMULPLAIN} operations (2 for vector-matrix multiplication and 1 for average computation due to multiplying by the reciprocal of number of words), $\lceil \frac{m}{t} \rceil$ \textsf{HADD} operations and $\log_2{N}$ \textsf{ROTATE} operations are the total number of homomorphic operations evaluated by the server. We note that the noise accumulated in ciphertexts due to homomorphic evaluations is mainly affected by \textsf{HMULPLAIN} and \textsf{ROTATE}. Concrete parameters for CKKS and \privft{} models are provided in Section~\ref{Performance:Evaluation}.

To achieve an efficient implementation, we parallelize vector-matrix operations and launch independent CPU threads to execute them on the available processing units. For instance, the computation of the $\mathbf{v}\cdot \mathbf{H}$ is executed by multiple threads such that each thread handles one column in matrix $\mathbf{H}$. Similarly, the dot-product $\mathbf{h}\cdot \mathbf{O}$ is executed by multiple threads, each handling a specific row in $\mathbf{O}$. In terms of memory utilization, the inference task can fit entirely in either CPU or GPU memory as it runs at small FHE parameters and includes a small number of ciphertexts.

\subsubsection{Homomorphic Training Implementation}

As mentioned previously, we had to adapt the training procedure in \fasttext{} to use minibatch SGD instead of true SGD as shown in Algorithm~\ref{alg:training:minibatch}. First, suitable minibatch size and number of epochs are assumed\footnote{This can be done based on prior experience developed by training on similar tasks with public data.}. Besides, the model matrices are initialized randomly. A minibatch consists of a number of tokens read from the dataset. A token can be a word, bi-gram (two consecutive words), tri-gram, ... etc. The function read\_batch reads the input training dataset one line at a time until number of tokens reaches $\beta$. The predict procedure in line 6 is implemented as in homomorphic inference except that the weight matrices here are encrypted, thus \textsf{HMUL} is used instead of \textsf{HMULPLAIN}. This step requires 2 levels due to matrix-vector multiplication to compute the hidden embedding vector and the multiplication by inverse of number of tokens in the read example. Generating the class scores output vector requires 2 \textsf{HMUL} operations, 1 \textsf{HMULPLAIN} for average computation, $\lceil \frac{m}{t} \rceil$ \textsf{HADD} operations and $\log_2{N}$ \textsf{ROTATE} operations. Next, the \textsf{softmax} function is applied to the output vector requiring 1 \textsf{HUML}. This is followed by calculating the error function between each of the scores in the output vector and the truth label. Finally, the gradient is accumulated until all the examples in minibatch are evaluated where the weight matrices are updated according to the general formula: $\text{model}_{\tau+1} = \text{model}_{\tau} - \eta \cdot \text{gradient}$. Note that this task also includes a number of mask and shift operations to extract columns/rows from the weight matrices for the update equations. To estimate the multiplicative depth of this task, an empirical approach is best used to count for the noise generated from rotations.

For the training task, the main source of parallelism is at the minibatch evaluation level. We launch multiple CPU threads, each handling a number of examples ($\frac{|\text{minibatch}|}{\text{num\_threads}}$) on one GPU card in a multi-GPU system. As both matrices $\mathbf{H}$ and $\mathbf{O}$ are encrypted, their size is significantly enlarged. Hence, we were not able to fit them inside the GPU local memory. Instead, the matrices are divided, vertically for $\mathbf{H}$ and horizontally for $\mathbf{O}$, and ported from CPU memory to GPU memory and back forth.


\begin{algorithm}[!ht]
         \caption{\textsf{GDMiniBatchTraining}} \label{alg:training:minibatch}
         \begin{algorithmic}[1]
         \Statex{Let T denote the total number of tokens in the dataset, where a token refers to a word, bi-gram or tri-gram.}
         \Statex{\textbf{Input}: training dataset (train\_data), num\_epochs, learning rate $\eta$, and $\beta$ the number of tokens in a minibatch}
         \Statex{\textbf{Output}: a trained model (model)}
         \State{model = initialize()}
         \State{i = 0}
         \While{i < (num\_epochs * T)}
            \State{batch = read\_batch($\beta$)}
            \State{labels = ground\_truth(batch)}
            \State{predictions = predict(batch, model)}
            \State{error = calculate\_error(labels, predictions)}
            \State{gradient = calculate\_gradient(error)}
            \State{model = update\_model(model, gradient)}
            \State{i = i + $\beta$}
         \EndWhile
         \State \textbf{return} model
         \end{algorithmic}
 \end{algorithm}




\section{Performance Evaluation}\label{Performance:Evaluation}
In this section, we evaluate the performance of \privft{}. We start by describing our testing methodology and platform configuration. Next, we present the datasets used for inference and training. Lastly, we present our results and discuss the key observations.

\subsection{Methodology}
The performance of \privft{} is evaluated using three experiments. Firstly, we evaluate the performance of our GPU implementation of the CKKS scheme. We design a timing experiment to measure the latency of the main cryptographic primitives used in the homomorphic evaluation of \privft{} inference and training tasks. Moreover, we compare the performance of our GPU CKKS implementation with a CPU implementation provided in Microsoft SEAL v3.3.0. We provide latency figures for different parameter settings. The main purpose of this experiment is to show the speedup one can obtain from GPU acceleration.

Secondly, we evaluate the performance of the homomorphic inference in \privft{}. We design a timing experiment to measure the latency of the inference. We use 6 different publicly available datasets in our evaluation. The performance is evaluated in both CPU and GPU implementations for comparison. Lastly, we evaluate the performance of homomorphic training in \privft{}. This task is both more complex and highly computational extensive. Due to limited access to the borrowed execution multi-GPU platform, we only evaluate it on a single dataset. Figures reported in experiments 1 and 2 are averaged over 100 runs. On the other hand, experiment 3 is only run once.

We are also interested in the communication cost between the client and server in both tasks (homomorphic inference and training). The evaluation part shows how to calculate the number and size of exchanged messages.

\subsection{Platform Configuration}

Experiments were performed on two different machines whose configurations can be found in Table~\ref{tab:machine:config}. The first is a commodity server with 2 CPUs each with 26 cores with hyper-threading enabled. In total, one may assume there are 104 CPU cores. The other platform is an NVIDIA DGX-1 multi-GPU system with 8 V100 GPU cards.

\begin{table}[!ht]
 	\small
	\caption {Hardware configuration of the testbed servers. PU stands for processing units.} \label{tab:machine:config}
	\centering
	\begin{tabular}{@{}l l| l@{}}
		\toprule
		{Specification} & \multicolumn{1}{c}{CPU}	&  \multicolumn{1}{c}{GPU}
		\\
		\midrule        
		\text{Model}                 & {{Intel R Xeon E5-2620}} & DGX-1 \\
		\text{GPU Model}             & \multicolumn{1}{c|}{-} & V100 \\
		\text{Compute Capability}    & \multicolumn{1}{c|}{-}  & 7.0 \\
		\text{\# of PUs}             & 54               & 8\\ 
		\text{\# of Cores (total)} 	 & 104             & $8 \times 5120$\\
		\text{Core Frequency}   	 & 2.40 GHz        & 1.380 GHz \\ 
		\text{Memory Capacity}  	 & 180 GB      & $8 \times 16$ GB \\
		\text{CUDA Version}		 & \multicolumn{1}{c|}{-}	& CUDA 9.0\\
		\text{Operating System}	& ArchLinux 4.19.32-1	& Ubuntu 4.4.0-142\\
		\text{Compiler}	& gcc 8.2.1	& gcc 4.9.4\\
		\bottomrule
	\end{tabular}
\end{table}


\subsection{Datasets}
We evaluate our method on various datasets covering three common text classification tasks: sentiment analysis, spam detection and topic classification. These datasets and their statistics are shown below:
\begin{itemize}
    \item IMDB: 50,000 movie reviews with binary sentiments (positive, negative) evenly distributed and so is the train/test split (thus 25,000 train examples and 25,000 test examples)
    
    \item Yelp: for each sentiment (positive, negative) there are 280,000 training samples and 19,000 testing samples. In total there are 560,000 training samples and 38,000 testing samples.
    
    \item AGNews: the AG's news topic classification dataset consists of 4 classes. Each class contains 30,000 training samples and 1,900 testing samples. The total number of training and testing samples are 120,000 and 7,600, respectively.
    
    \item DBPedia: the DBpedia ontology classification dataset is constructed by picking 14 non-overlapping classes from DBpedia 2014. For each class, there are 40,000 training samples and 5,000 testing samples. The total size of the training dataset is 560,000 examples and testing dataset is 70,000 examples.
    
    \item Youtube Spam Collection: the data set is collected from UCI Machine Learning Repository. It is a public set of comments collected for spam research. It has five datasets composed by 1,956 real messages extracted from five videos that were among the 10 most viewed on the collection period.
    
    \item Enron Email Dataset: total of 35,715 emails, of which we use 80\%/20\% split for train and test.
\end{itemize}

Note that these datasets vary in the number of classes to be predicted. For instance, YouTube, Enron Email, IMDB and Yelp are binary classification problems that include only 2 classes. On the other hand, AGNews and DBPedia are multiclass problems with 4 and 14 classes, respectively.

\subsection{CKKS Parameter Selection}
The \textit{levelled} CKKS includes a number of parameters that must be set appropriately to ensure both correctness and sufficient security level. First, the standard deviation of the discrete Gaussian distribution $\sigma$ is set to 3.2 following the recommendations of the FHE standard~\cite{HomomorphicEncryptionSecurityStandard}. 

There are three problem-dependent parameters: $L$, $\rho$ and $q_L$. In \privft{}, the inference task requires $L = 5$ to cater for three subsequent multiplications and vector rotations in the dot products. The training task, on the other hand, requires larger depth $L = 46$ levels to cater for the feed-forward phase and computing the loss function all multiplied by the number of epochs (2 in our case). The CKKS computation precision $\rho$ has a limited range practically $(2^{20}, \cdots,  2^{60})$ and can be chosen experimentally. Finally, the size of $q_L$ in bits can be estimated heuristically as $|q_L| = L\times \rho$ bits. The final parameter that needs to be set is the ring dimension $N$ which affects both performance and security level $\lambda$. $N$, which is a power of 2 number, has a very limited range $\{ 2^{10}, \cdots, 2^{17} \}$ in practical implementations. For each $N$, one can use the LWE estimator~\cite{albrecht2015concrete} to estimate the security level achieved. According to NIST recommendations~\cite{barker2012recommendation}, the minimum security level recommended for today's computing capacity is $80$-bit.

\subsection{CKKS Micro-Benchmarks and Comparison}
In this section, we compare the performance of our GPU implementation of CKKS with Microsoft SEAL~\cite{SEAL} version 3.2, which implements an RNS variant of the CKKS scheme in C++. 

Table~\ref{tab:bench:seal} shows the latency of a number of CKKS arithmetic primitives in both SEAL and our GPU-accelerated implementation. We only show the critical arithmetic operations that are heavily used in homomorphic computation. Other less critical operations such as \textsf{KEYGEN}, \textsf{ENC}, \textsf{DEC}, \textsf{ENCODE}, and \textsf{DECODE} are only used at system initialization and input/output stages. We note that the parameter sets $(\log_2{N},\log_2{q}) = (13,200)$ and $(\log_2{N},\log_2{q}) = (16,2300)$ are used in \privft{} homomorphic inference and training tasks, respectively. We include them here to give more insights on the complexity of the core operations used to implement each task.

It can be seen that GPU-CKKS outperforms SEAL-CKKS in all primitives with various parameter settings. Speedup factors ranging from 1 to 2 orders of magnitude are observed. Of particular interest are \textsf{HMULPLAIN} and \textsf{HMUL} which are heavily used in inference and training, respectively. In particular, \textsf{HMULPLAIN} showed 33.78$\times$ to 244.62$\times$ improvement over SEAL whereas \textsf{HMUL} shows 26.88$\times$ to 48.22$\times$.
\begin{table*}[htbp]
\scriptsize
  \centering
  \caption{Latency in (milliseconds) and speedup factors of core CKKS homomorphic operations in Microsoft SEAL (CPU) and our GPU-accelerated implementation. LHW and HHW denote low and high hamming weight of the rotation step. REL refers to relinearization. G, C, and S refer to GPU, CPU and Speedup.}
    \begin{tabular}{@{} lrrr|rrr|rrr|rrr|rrr@{}}
    \toprule
    & \multicolumn{15}{c}{{\normalsize{$(\log_2{N}, \log_2{q},\lambda)$}}} \\
    \cmidrule{2-16}
    \multicolumn{1}{c}{} & \multicolumn{3}{c|}{(13,200,140)} & \multicolumn{3}{c|}{(14,360,159)} & \multicolumn{3}{c|}{(15,600,196)} & \multicolumn{3}{c|}{(16,1770,128)} & \multicolumn{3}{c}{(16,2300,98)} \\
\cmidrule{2-16}    \multicolumn{1}{c}{{\small{Procedure}}} & \multicolumn{1}{c}{G} & \multicolumn{1}{c}{C} & \multicolumn{1}{c|}{S$\times$} & \multicolumn{1}{c}{G} & \multicolumn{1}{c}{C} & \multicolumn{1}{c|}{S$\times$} & \multicolumn{1}{c}{G} & \multicolumn{1}{c}{C} & \multicolumn{1}{c|}{S$\times$} & \multicolumn{1}{c}{G} & \multicolumn{1}{c}{C} & \multicolumn{1}{c|}{S$\times$} & \multicolumn{1}{c}{G} & \multicolumn{1}{c}{C} & \multicolumn{1}{c}{S$\times$} \\
\cmidrule{1-16}    
    \multicolumn{1}{@{}l|}{HADDPLAIN} & \textbf{0.03} & 0.39  & 11.91 & \textbf{0.04} & 0.83  & 21.87 & \textbf{0.05} & 3.08  & 57.11 & \textbf{0.18} & 28.48 & 154.76 & \textbf{0.30} & 32.80 & 109.70 \\[.075cm]
    \multicolumn{1}{@{}l|}{HMUL+REL} & \textbf{0.40} & 10.83 & 26.88 & \textbf{0.74} & 23.07 & 31.10 & \textbf{2.34} & 107.08 & 45.80 & \textbf{33.58} & 1619.15 & 48.22 & \textbf{55.88} & 2519.85 & 45.09 \\[.075cm]
    \multicolumn{1}{@{}l|}{HMULPLAIN} & \textbf{0.02} & 0.61  & 33.78 & \textbf{0.02} & 1.27  & 66.74 & \textbf{0.04} & 4.10  & 117.14 & \textbf{0.14} & 26.93 & 198.00 & \textbf{0.17} & 41.59 & 244.62 \\[.075cm]
    \multicolumn{1}{@{}l|}{RESCALE} & \textbf{0.12} & 5.60  & 47.47 & \textbf{0.14} & 11.85 & 84.61 & \textbf{0.27} & 41.29 & 150.69 & \textbf{1.28} & 257.46 & 200.67 & \textbf{1.63} & 319.19 & 195.58 \\[.075cm]
    \multicolumn{1}{@{}l|}{ROTATE\_LHW} & \textbf{0.62} & 14.67 & 23.51 & \textbf{0.88} & 37.55 & 42.62 & \textbf{2.55} & 186.24 & 72.95 & \textbf{39.91} & 2018.00 & 50.56 & \textbf{55.64} & 2964.35 & 53.28 \\[.075cm]
    \multicolumn{1}{@{}l|}{ROTATE\_HHW} & \textbf{3.74} & 85.37 & 22.83 & \textbf{6.09} & 258.07 & 42.35 & \textbf{17.82} & 1292.73 & 72.55 & \textbf{324.90} & 16022.90 & 49.32 & \textbf{444.79} & 24144.61 & 54.28 \\
    \bottomrule
    \end{tabular}%
  \label{tab:bench:seal}%
\end{table*}%

The rotation procedure shows varied latency figures depending on the Hamming Weight (HW) of the number of desired rotations ($\pi$). Note that for every bit (that is set, i.e., has a value of 1) in the binary representation of $\pi$, a Galois automorphism is applied to the ciphertext, such that $X \rightarrow X^\kappa$, where $\kappa$ is the Galois element that corresponds to a cyclic row rotations of magnitude $\pi$ to the left/right depending on the sign of $\pi$. For every mapping, a key switching operation is invoked with a corresponding rotation key that is pre-generated at the initialization time. As we can rotate in both directions, one needs to perform $\frac{\log_2{N}}{2}$ such rotations at most when the HW of $\pi$ is the maximum. This justifies the variations in the rotation procedures latency. Note that one can pre-generate $N$ rotation keys (for both directions) and perform any desired rotation by applying a single Galois automorphism (at the cost of \textsf{ROTATE\_LHW}), but that may not be practical as the rotation keys are usually huge in size. For instance, in our RNS implementation of the CKKS scheme, one rotation key contains $l$ ciphertexts, where $l$ is the number of primes factors in the ciphertext coefficient modulus $q$. We remark that we could not keep the rotation keys in GPU memory at high parameters $(\log_2{N}=16)$ due to their huge size. The keys are instead stored in CPU memory and migrated to GPU when needed. This explains the high latency for rotations. For further details on applying Galois theory for FHE ciphertext rotations, the reader is referred to~\cite{gentry2012fully}.

It should be remarked that we had to adapt SEAL to support ring dimension $N=2^{16}$ as it only supports up to $2^{15}$ ring dimensions by default.

\subsection{\privft{} Micro-Benchmarks}

In the following paragraphs, we present the performance of inference and training in \privft{}.

\subsubsection{\privft{} Inference Task }

In order to run this task, we first generate the models using \fasttext{}. The models are generated for all 6 tasks. Note that training for inference is done in plaintext domain. To evaluate the quality of the models generated by \fasttext{}, we compare with recent approaches as shown in Table~\ref{tab:privft:inference:accuracy}. In this experiment, we use $m = 500,000$ to limit the vocabulary size, and set $n = 50$. Further, we include both bi-gram and tri-gram terms to the input by augmenting the text sequence to additionally contain these terms (thus creating a longer sequence). Training (to generate the models for \privft{} inference) was done using stochastic gradient descent (i.e. minibatch size $= 1$). Note that in contrast to ULMFit, BERT ITPT, and XLNet which all include a pre-training phase on some large out-of-domain datasets, PrivFT only uses the training dataset given in each task. We emphasize that our goal is not to advance the state-of-the-art in text classification, but rather to outline a competitive yet practical system on encrypted data.

\begin{table}[!ht]
  \centering
  \caption{Accuracy (\%) of our unencrypted \fasttext{} model against the current state of the art (\textsf{ULMFit}~\cite{howard2018universal}). 
    }
    \begin{tabular}{lrrrr}
    \toprule
     \multicolumn{1}{c}{\textbf{Model}} & \multicolumn{4}{c}{\textbf{Datasets}} \\
\cmidrule{2-5}          & \multicolumn{1}{c}{\textbf{Yelp }} & \multicolumn{1}{c}{\textbf{AG}} & \multicolumn{1}{c}{\textbf{IMDB }} & \multicolumn{1}{c}{\textbf{DBPedia}} \\
    \midrule
    \textbf{XLNet}~\cite{yang2019xlnet} & 98.45 & 95.51 & 96.21 & 99.38\\
    \textbf{BERT ITPT}~\cite{sun2019fine} & 98.19 & 95.34 & 95.79 & 99.39\\
    \textbf{ULMFit}~\cite{howard2018universal} & 97.84 & 94.99 & 95.40 & 99.20\\
    \textbf{PrivFT} & 96.06 & 92.54 & 91.49 & 98.80 \\
    \bottomrule
    \end{tabular}%
  \label{tab:privft:inference:accuracy}%
\end{table}%

We implemented the inference task of \privft{} in both SEAL-CKKS and our GPU-accelerated CKKS. The CKKS parameters used in this experiment are: $(N, \log_2{q_{L}}, \rho) = (8192, 200, 40)$, providing sufficient security level $\lambda = 140$ bit. Table~\ref{tab:privft:inference:results} shows the latency (in seconds) of evaluating the inference task for one example from each dataset in both SEAL and our GPU implementation (here we also include the Youtube Spam Collection and Enron Email Dataset for benchmarking purposes). Our GPU implementation provides a quite reasonable run time ($<0.66$ seconds) for practical applications with 12$\times$ speedup compared to CPU. We emphasize that the prediction accuracy of \privft{} on encrypted data is the same as \fasttext{} accuracy on unencrypted data as shown in Table~\ref{tab:privft:inference:accuracy}.

~\\
\textbf{Message Size~~} The client is required to communicate to the server $\lceil \frac{m}{t} \rceil$ ciphertexts, i.e. $\lceil \frac{500000}{4096} \rceil = 123$ ciphertexts. The ciphertext size in bits can be estimated as $2*N*\log_2{q_L}$. Therefore, the total message size from the client to the server is 384.375 MB. After homomorphic inference, the server sends back to the client only 1 ciphertext (6.25 MB) given that the number of classes is $\leq N/2$, which is the case in all datasets used in our evaluation.

\begin{table*}[!ht]
  \centering
  \caption{Latency (in seconds) of evaluating \privft{} inference task with Microsoft SEAL 3.2 and our GPU-accelerated implementation for different datasets.}
    \begin{tabular}{lrrrrrr}
    \toprule
     \textbf{Scheme} & \multicolumn{6}{c}{\textbf{Datasets}} \\
\cmidrule{2-7}          & \multicolumn{1}{c}{\textbf{Youtube Spam}} & \multicolumn{1}{c}{\textbf{Enron Email}} & \multicolumn{1}{c}{\textbf{Yelp }} & \multicolumn{1}{c}{\textbf{AGNews}} & \multicolumn{1}{c}{\textbf{IMDB }} & \multicolumn{1}{c}{\textbf{DBPedia}} \\
    \midrule
    \textbf{SEAL (CPU)} & 3.84 & 7.95 & 7.88 & 7.88 & 7.90 & 7.74 \\
    \textbf{Ours (GPU)} & \textbf{0.23} & \textbf{0.63} & \textbf{0.65} & \textbf{0.66} & \textbf{0.65} & \textbf{0.63} \\
    \midrule
    \textbf{Speedup} & 16.69$\times$ & 12.62$\times$ & 12.08$\times$ & 12.00$\times$ & 12.17$\times$ & 12.25$\times$ \\
    \bottomrule
    \end{tabular}%
  \label{tab:privft:inference:results}%
\end{table*}%
\subsubsection{\privft{} Training Task }

The training task is more computationally intensive compared to the inference task. To attain sufficient multiplicative depth and security level, we had to set the CKKS parameters to $(N, \log_2{q_{L}}, \rho) = (65536, 2300, 50)$. The number of epochs and batch size $\beta$ were set to 2 epochs and 1007500 tokens, respectively. This large batch size limits the number of model aggregates resulting in a training circuit with a shallow multiplicative depth. With these settings, the number of minibatches evaluated in this task was 5. Each minibatch requires 9 levels of multiplicative depth. Thus, we require at least $9 \cdot 5 = 45$ levels to ensure that decryption works successfully. We remark that we needed 1 extra level to get the same accuracy as that of training in plaintext domain. Note that $\log_2{q_{L}}/ \rho = 46$, which is the multiplicative depth required by this task. 

We implemented and ran the training task for the YouTube Spam Collection dataset with both CKKS implementations (CPU and GPU). Training in SEAL on CPU took 266.4 hours (or 11.1 days) on 104 cores. Under these large parameters, the memory requirements of the training task could not be met by our GPU system. Therefore, we had to utilize tiling techniques and decompose the computation into subtasks that fit in system memory. Note that \privft{} training requires roughly 120 GB RAM. Training on 8 GPUs (with 8 CPU threads) took 120.96 hours (or 5.04 days) achieving 2.2$\times$ speedup compared to CPU. Note that speedup for training is relatively smaller than that of inference due to the computation tiling and data migration between CPU memory and GPU memory. We remark that the training procedure is embarrassingly parallel that can benefit from additional dedicated hardware resources. We expect that this task to show close to optimal scalability as the number of computing resources is increased. We also remark that the accuracy of the generated model (by SEAL and our library) was the same as that generated by \fasttext{} using gradient descent with large minibatch, that is 86.3\%.

~\\
\textbf{Message Size~~} The client needs to encrypt each record in the dataset using $\lceil \frac{m}{t} \rceil$ ciphertexts. The ciphertext size in this task is 287.5 MB, i.e., total message size will be $ \text{num\_records} \cdot 287.5$ MB, where $\text{num\_records}$ is the number of records in the training dataset. After training is done, the server sends back to the client $n(\lceil \frac{m}{t} \rceil + c)$ ciphertexts.

~\\
\textbf{Discussion~~}
Although we showed that \privft{} is able to perform the training task for a non-trivial NLP text classification problem, we remark that there are a number of challenges that need to be addressed before homomorphic training can become a mainstream service. First, there are a number of hyperparameters that need to be fine-tuned to attain effective models. Some of these parameters are the learning rate, number of epochs and minibatch size. The best approach to find these parameters is through trial and error or following a heuristic search method. Both (or at least the former) are impractical in FHE as the server cannot evaluate the validity of the model. A cooperative approach with the client can be adopted but that may not be desirable.

Another challenge related to this task is the multiplicative depth of the training circuit. As the number of epochs or the degree of the loss function increases, the multiplicative depth is enlarged proportionally which might incur ultra-large FHE parameters or bootstrapping to support the required depth. Moreover, the fixed-point approach employed in CKKS generates higher precision loss compared to the standard floating-point approach used in plaintext~\cite{cheon2017homomorphic}. This means that with deeper circuits, one has to start with a large CKKS computation precision $\rho$, that may not be supported by existing libraries and require special treatment.

Lastly, our solution requires ultra large minibatch size and low number of epochs which may not work out for all tasks. However, some recent works showed that SGD with large minibatch size can still provide effective training~\cite{goyal2017accurate}. Moreover, other works showed that second-order optimization methods such as Kronecker-Factored Approximate Curvature (K-FAC), which requires very large minibatch size, provide competitive results with SGD on the ImageNet dataset~\cite{Osawa_2019_CVPR}. It is an interesting problem to investigate whether K-FAC can be implemented with FHE.

\section{Conclusions}
\label{sec:conclusion}
In this work, we proposed \textbf{\privft{}}: \textbf{Priv}ate and \textbf{F}ast \textbf{T}ext classification solution on encrypted data using FHE. The main tasks of \privft{} were: 1) to perform inference on encrypted data using a pre-learned plaintext model and 2) training an effective model on encrypted data to generate an encrypted model. \privft{} can be used to implement several text classification applications such as sentiment analysis, spam detection, topic classification and document classification without compromising the privacy of the input data. 
We provided an efficient GPU implementation of an RNS variant of the CKKS FHE scheme and compared its performance with an existing CPU implementation provided in Microsoft SEAL. Experiments showed that 1 to 2 orders of magnitude speedup can be achieved. We implemented \privft{} in both CPU and GPU libraries and showed that \privft{} requires less than 0.66 seconds (on GPU) per inference on various datasets. We also showed how training a model on encrypted data can be done in \privft{}. Unfortunately, homomorphic training in \privft{} was more computationally intensive and took 11.1 days and 5.04 days on CPU and GPU, respectively.





\section*{Availability}

The source code of \privft{} in Microsoft SEAL will be made publicly available when the paper is published.

\bibliographystyle{plain}
\bibliography{biblio}

\end{document}